\documentclass[journal]{IEEEtran}

\usepackage{cite}
\usepackage{amsmath,amssymb,amsfonts}
\usepackage{algorithm}
\usepackage{algorithmicx}
\usepackage{algpseudocode}
\usepackage{graphicx}
\usepackage{enumitem}
\usepackage{textcomp}
\usepackage{xcolor}
\usepackage{caption}
\usepackage{subcaption}
\usepackage{float}
\usepackage{multirow}
\usepackage{xspace}
\usepackage{booktabs}
\usepackage{url}
\usepackage{amssymb}
\usepackage{amsmath}
\usepackage[most]{tcolorbox}
\usepackage{multirow}
\usepackage[colorlinks,linkcolor=black, urlcolor=blue]{hyperref}

\newcommand{\sz}[1]{\textcolor{magenta}{{#1}}}

\newcommand{\tool}{{DTrans}\xspace}

\newcommand{\sml}{{$M_{small}$}\xspace}
\newcommand{\medium}{{$M_{medium}$}\xspace}
\newcommand{\gpj}{{GitProjs}\xspace}
\newcommand{\parabf}[1]{\noindent \textbf{#1}}

\newcommand{\rltrans}{Transformer\textsubscript{relative}}

\begin{document}

\title{Dynamically Relative Position Encoding-Based Transformer for Automatic Code Edit}

\author{Shiyi~Qi,
        Yaoxian~Li,
        Cuiyun~Gao, Xiaohong Su,
        Shuzheng~Gao, Zibin Zheng, and Chuanyi Liu
\thanks{S.Qi, Y.Li, C. Gao, X. Su, S. Gao, C. Liu were with Harbin Institute of Technology, China (e-mail: syqi981125@163.com, yaoxian0803@icloud.com, gaocuiyun@hit.edu.cn, sxh@hit.edu.cn, szgao98@gmail.com, liuchuanyi@hit.edu.cn).}
\thanks{C. Gao and C. Liu were also with Peng Cheng Laboratory and Guangdong Provincial Key Laboratory of Novel Security Intelligence Technologies, China.}
\thanks{Z. Zheng was with Sun Yat-sen University, China (email: zibinzheng2@yeah.net).}
\thanks{C. Gao and C. Liu are the corresponding authors.}
\thanks{Manuscript received April 19, 2005; revised August 26, 2015.}
}
%
%

\markboth{Journal of \LaTeX\ Class Files,~Vol.~14, No.~8, August~2015}%
{Shell \MakeLowercase{\textit{et al.}}: Bare Demo of IEEEtran.cls for IEEE Journals}

\maketitle


\begin{abstract}

Adapting Deep Learning (DL) techniques to automate non-trivial coding activities, such as code documentation and defect detection, has been intensively studied recently. Learning to predict code changes is one of the popular and essential investigations.
Prior studies have shown that DL techniques such as Neural Machine Translation (NMT) can benefit meaningful code changes, including bug fixing and code refactoring.
However, NMT models may encounter bottleneck when modeling long sequences, thus are limited in accurately predicting code changes. In this work, we design a Transformer-based approach, considering that Transformer has proven effective in capturing long-term dependencies. Specifically, we propose a novel model named DTrans. For better incorporating the local structure of code, i.e., statement-level information in this paper, DTrans is designed with dynamically relative position encoding in the multi-head attention of Transformer. Experiments on benchmark datasets demonstrate that DTrans can more accurately generate patches than the state-of-the-art methods, increasing the performance by at least 5.45\%-46.57\% in terms of the exact match metric on different datasets. Moreover, DTrans can locate the lines to change with 1.75\%-24.21\% higher accuracy than the existing methods.



\end{abstract}


%
\IEEEpeerreviewmaketitle

\section{Introduction}\label{sec:intro}

Deep Learning (DL) techniques have been adapted to solve many traditional software engineering problems and tasks recently \cite{ferreira2021software,yang2020survey}, e.g., fault localization\cite{eniser2019deepfault,li2019deepfl,wardat2021deeplocalize}, automatic program repair\cite{lutellier2019encore,li2020dlfix,gupta2017deepfix}, code summarization \cite{ahmad2020transformer,leclair2020improved,wan2018improving}, code prediction \cite{kim2021code, huang2020commtpst}, and defect prediction\cite{hasanpour2020software,chen2020software,wang2018deep}. Among these fields, learning from code for code change prediction draws more and more research investigations\cite{islam2019comprehensive,wen2018well}. Precisely 
editing code can significantly facilitate the software maintenance process for developers \cite{tufano2019learning,chakraborty2020codit}.

\begin{figure*}[htbp]
    \centering
    \begin{subfigure}[b]{0.48\textwidth}
        \includegraphics[width=\textwidth]{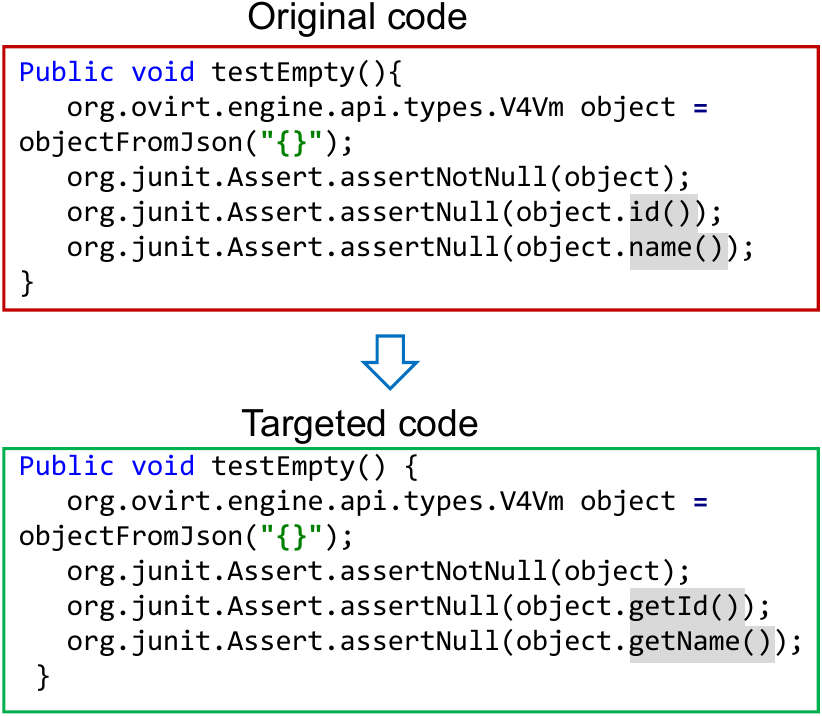}
        \caption{Example 1.}
      \end{subfigure}
      \hfill
      \begin{subfigure}[b]{0.48\textwidth}
        \includegraphics[width=\textwidth]{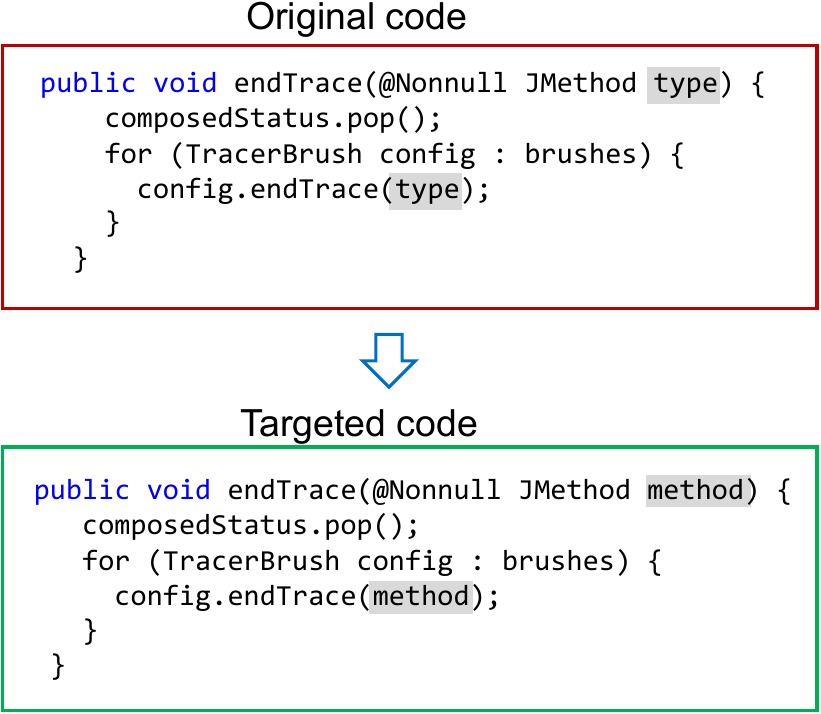}
        \caption{Example 2.}
        \label{fig:layer}
      \end{subfigure}
    \caption{Examples for illustrating the code edit task. The grey blocks highlight the changed parts in the code.}
    \label{fig:example}
\end{figure*}

In the process of program development and maintenance, developers usually need to modify the source code for various reasons, including
program repair~\cite{lutellier2020coconut,jiang2021cure}, code refactoring~\cite{tansey2008annotation,meng2015does,ge2012reconciling}and API-related changes~\cite{nguyen2010graph,nguyen2016api},  etc. Such behavior is known as ``code edit'' or ``code change''~\cite{tufano2019learning,chakraborty2020codit}.
Prior research ~\cite{tufano2019learning,chakraborty2020codit,tufano2018empirical,chen2019sequencer} discovers that code edits
generally follow repetitive edit patterns and can be employed to automatically generate targeted code based on original code.
Figure~\ref{fig:example} shows two examples for illustrating the code edit task.
In the original code of Figure~\ref{fig:example} (a), the method \texttt{testEmpty} needs to return the object's ID and name. However, the functions \texttt{id()} and \texttt{name()} do not exist for the object, which leads to a program bug. 
The correct code edit operation is to generate a correct patch for fixing the bug, i.e, changing to the corresponding correct functions \texttt{getId()} and \texttt{getName()}, respectively.
For the example in Figure~\ref{fig:example}(b), the parameter name is changed from \texttt{type} to \texttt{method} for enhancing the readability of the code. The code edit task aims at generating the edited code given the original code\cite{tufano2019learning}.
Due to the complex code edit patterns, automatically identifying the lines for editing and producing accurate edits are challenging.


In recent years, deep learning has made great progress and been applied to many code-related tasks \cite{DBLP:journals/corr/abs-2009-08366,DBLP:conf/kbse/TianLKKLKB20}. The large software engineering datasets, such as GitHub which includes over 100 million repositories with over 200 million merged pull requests (PRs) and 2 billion commits\cite{tufano2019learning}, provide us with sufficient source code for training DL models.
Prior studies~\cite{chen2019sequencer,tufano2019empirical,tufano2019learning} have shown that DL techniques such as Neural Machine Translation (NMT)~\cite{bahdanau2014neural} can automatically apply developers' pull request code to generate meaningful code changes.
NMT models treat pull request code as a series of tokens or use the parsed tree structure as input, then creating an intermediate representation with an encoder network and decoding the representation into target sequence with a decoder network\cite{ahmad2020transformer,alon2018code2seq}. 
This mechanism makes NMT models can learn complex code change patterns between input and output sequences\cite{chakraborty2020codit}.
However, NMT models have proven ineffective in modeling long sequences~\cite{DBLP:conf/ijcnlp/LeMYM17}, thus may be limited in accurately editting code. 
Considering that Transformer~\cite{vaswani2017attention,DBLP:conf/iclr/RaePJHL20,DBLP:conf/nips/ZaheerGDAAOPRWY20} has shown more effective than NMT in modeling long sequences, it is more applicable for the task. But directly adopting Transformer still cannot well capture the structural dependencies between tokens \cite{ahmad2020transformer,wang2021code}. Thus, to mitigate the issue of NMT models and better capture the code dependencies, we propose a novel Transformer-based model, named as \tool.

Prior research~\cite{chakraborty2020codit} extracts the AST of original code for capturing the structural information. In this work, to alleviate the complexity caused by the AST extraction, we focus on exploiting the local structure, i.e., the statement-level information, which can be easily obtained without parsing. Besides, for the code editing task, the changes generally happen within several statements, indicating the importance of local structure\cite{chen2019sequencer,chakraborty2020codit}. 
Specifically, we propose a dynamically relative position encoding strategy to integrate the variational statement-level information. Different from the Transformer\cite{vaswani2017attention} and Transformer with relative position\cite{shaw2018self}, which represent position embedding by absolute position and relative position respectively, \tool conducts positional encoding guided by statement information.
 



To evaluate the performance of our proposed \tool model, we choose three pull request repositories utilized by the work~\cite{tufano2019learning} as benchmark datasets, including Android\cite{android}, Google Source\cite{google}, and Ovirt\cite{Ovirt}. 
Besides, we also involve the 2.3 million 121,895 pair-wise code changes from GitHub open-source projects\cite{tufano2019empirical}.
During evaluation, we group project datasets to two levels, i.e., small and medium levels, according to the token numbers of original code following prior studies\cite{cho2014learning,tufano2019empirical,chen2019sequencer}.
Experiments demonstrate that \tool accurately predicts more code edits in both small-level and medium-level projects, increasing the performance of the best baseline \cite{chen2019sequencer} by at least 5.45\% and 25.76\% in terms of the exact match metric, respectively. Moreover, \tool successfully locates the lines to change with 1.75\%-24.21\% higher accuracy than the best baseline.

Overall, we make the following contributions:

\begin{itemize}
    \item A novel Transformer-based approach is proposed to incorporate dynamically relative position encoding strategy in the multi-head attention of Transformer, which explicitly incorporates the statement-level syntactic information for better capturing the local structure of code.
    \item  We evaluate our approach on benchmark datasets, and the results demonstrate the effectiveness of \tool in predicting accurate code changes.
\end{itemize}


\textbf{Paper structure.} We introduce the background in Section~\ref{sec:back}. The proposed approach is illustrated in Section~\ref{sec:method}. Experimental setup and results are depicted in Section~\ref{sec:setup} and Section~\ref{sec:results}, respectively. We show some cases in Section~\ref{sec:case}. The threats to validity and related work are introduced in Section~\ref{sec:threat} and Section~\ref{sec:literature}, respectively. We conclude our work in Section~\ref{sec:con}.

\section{Background}\label{sec:back}
In this section, we first formulate the code change prediction task and then introduce the basic approach - Transformer. 

\subsection{Deep learning (DL) in Code Change Prediction}

DL-based techniques aim at learning the mapping relations between the original code and the target code by training, and generating edited code for facilitating software development\cite{bhatia2018neuro,chen2019sequencer,dinella2020hoppity}. 
Programming languages can be treated as sequences of code tokens. Therefore, the problem of code change prediction can be tackled as a neural machine translation problem\cite{tufano2019empirical,chakraborty2020codit}, that is, to ``translate'' from a sequence of code tokens (the original code) to another sequence of code tokens (the target code).  

We take a a sequence of original code $\mathbf{O}$ as an example, and let
\[
\mathbf{O}=(o_1,o_2,...,o_i,...o_m),
\]
where each $o_i$ is the $i$-th token in the code. Each input sequence $\mathbf{O}$ corresponds to a target code $\mathbf{C}$, denoted as:
\[\mathbf{C}=(c_1,c_2,...,c_n),\]
where $m$ and $n$ indicate the lengths of the original and target sequences, respectively. 
Our goal is to learn the conditional distribution and generate changed code sequence by maximizing the conditional likelihood:
\[\mathbf{C}=\mathop{\arg\max}_\mathbf{C}P(\mathbf{C}|\mathbf{O}).\]
Finally, we achieve an optimized target sequence as the predicted code change. 


\begin{figure*}
    \centering
    \includegraphics[scale=0.7]{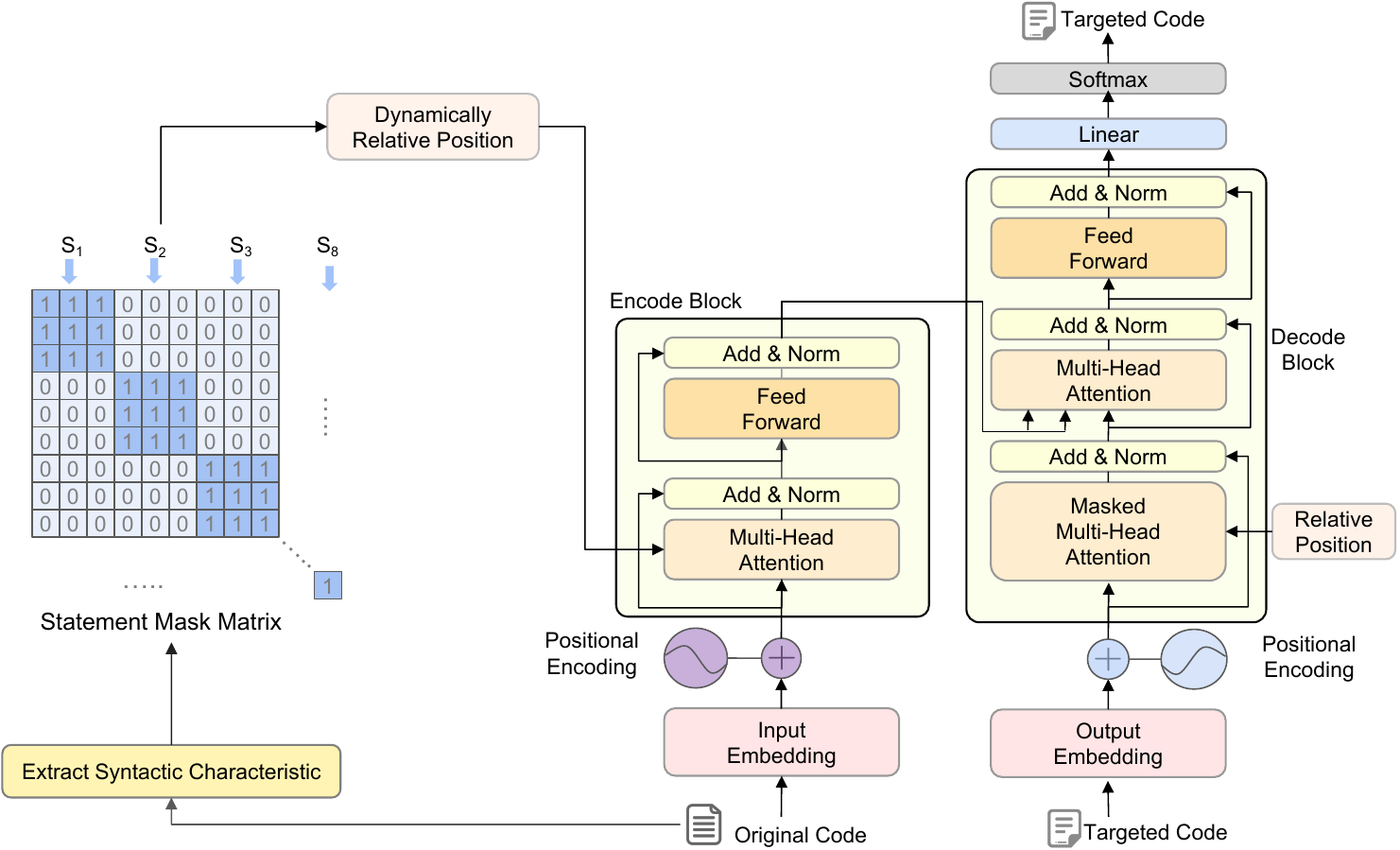}
    \caption{Framework of the proposed \tool. The statement mask matrix takes the code snippet shown in Figure~\ref{fig3} as example.}
    \label{fig:framework}
\end{figure*}

\subsection{Transformer}
\label{sec:transformer}
Transformer employs the typical encoder-decoder structure~\cite{cho2014learning}, and is composed of stacked Transformer blocks. Each block contains a multi-head self-attention sub-layer followed by a fully connected positional-wise feed-forward network sub-layer. The sub-layers are connected by residual connections~\cite{he2016deep} and layer normalization~\cite{ba2016layer}. In addition, Transformer augments the input features by adding a positional embedding since the self-attention mechanism lacks a natural way to encode the word position information. Transformer also applies pad masking to resolve the problem of variable input lengths and its decoder uses sequence masking in its self-attention to ensure that the predictions for the $i$-th position can only depend on the known outputs at positions less than $i$. We introduce the major components of Transformer, including multi-head self-attention, position-wise feed-forward networks, and basic blocks of Transformer in the following.

\subsubsection{Multi-Head Self-Attention} 
Multi-head self-attention involves multiple attention heads and performs self-attention mechanism on every head. One attention head obtains one representation space for the same text, and multi-head attention obtains multiple different representation spaces.
The self-attention mechanism can be described as mapping a query and a set of key-value pairs to an output, where the query, key, value, and output are all $d$-dimensional vectors. The output of each head is concatenated and results in the final output vector once again projected.

\par
\textbf{Scaled Dot-Product Attention.} The self attention used in Transformer is also known as scaled dot-product attention. Scaled dot-product attention aims to pay more attention to the important information of input sequence~\cite{vaswani2017attention}. It transposes the sequence of input vectors $\mathbf{X}=(x_1,x_2,...,x_n)$ into the sequence of output vectors $\mathbf{Z}=(z_1,z_2,...,z_n)$, where $x_i$, $z_i\in {R^{d_{model}}}$. When doing self attention, Transformer first projects the input vector $\mathbf{X}$ into three vectors: the query $Q$, key $K$ and value $V$ by trainable parameters $W^{Q}$, $W^{K}$, $W^{V}$. The attention weight is calculated using dot product and softmax function. The output vector is the weighted sum of the value vector:\par
\begin{equation}
    e_{ij}=\frac{(x_iW^Q)(x_jW^K)^T}{\sqrt{d}}, \label{eq:1}
\end{equation}
\begin{equation}
    \alpha_{ij}= \frac{exp\;e_{ij}}{\sum_{k=1}^{n} exp\;e_{ij}}, \label{eq:2}
\end{equation}
\begin{equation}
    z_i=\sum\limits_{j=1}^{n}\alpha_{ij}(x_{j}W^{V}), \label{eq:3}
\end{equation}
where $d$ is the dimension of each vector, and is used to scale the dot product.

\textbf{Multi-Head Attention.} Multi-head attention captures different context with multiple individual self-attention functions. This mechanism allows Transformer to jointly attend to information from different representation sub-spaces. Multi-head attention is computed after scaled dot-product attention:

\begin{equation}
    MultiHead(X)=Concat(head_1,...,head_h)W^O,  \label{eq:4}
\end{equation}

\begin{equation}
   head_i=Attention(XW_i^Q,XW_i^K,XW_i^V),  \label{eq:5}
\end{equation}

\noindent where $W^O$ indicates the learnable parameters and the parameters $W_i^Q$, $W_i^K$, $W_i^V$ are independent in each head.

\subsubsection{Position-wise Feed-Forward Networks}
 In addition to multi-head self-attention sub-layers, each block in encoder and decoder also contains a fully connected feed-forward network (FFN) sub-layer. FFN
transforms the current feature space into another space through non-linear mapping, aiming at learning a better representation of the input. The parameters of each position are shared. This FFN can be computed by two linear transformations and a ReLU activation function between them.
\begin{equation}
    FFN(x)=max(0,xW_1+b_1)W_2+b_2,      \label{eq:6}
\end{equation}

\noindent where $W_1$, $W_2$, $b_1$, and $b_2$ are learnable parameters.

\subsubsection{Basic Blocks of Transformer} 
Transformer is composed of stacked encoder-decoder blocks. 
Every block in the Transformer has a multi-head self-attention sub-layer and an FFN sub-layer. These sub-layers are connected by the residual connections (He et al.\cite{he2016deep}) and layer normalization (Ba et al.\cite{ba2016layer}). Different from the encoder block, the decoder block has another attention sub-layers that use the key and value matrices from the encoder instead of calculating them from the projection of input (\tool is a Transformer-based architecture, so the structure of encoder and decoder block also can refer to Fig. \ref{fig:framework}). Besides, the number of encoder-decoder blocks will affect the performance of Transformer, and more encoder-decoder block will increase the model size and require more time to train 



\section{Methodology}\label{sec:method}

In this section, we introduce the Transformer-based model \tool for automatic code change prediction. The overall architecture of \tool is shown in Figure~\ref{fig:framework}, following the general Transformer framework (as introduced in Section~\ref{sec:back}). In order to mitigate the out-of-vocabulary (OOV) problem, we first perform code abstraction following the prior work~\cite{ahmed2018compilation,tufano2019empirical}. Also, different from the vanilla Transformer, we propose a novel position encoding strategy, named \textit{dynamically relative position encoding}, to incorporate statement-level syntactic information into Transformer for better capturing the local structure of code. We elaborate on the code abstraction process, and the proposed dynamically relative position encoding in more details in the following.

\subsection{Code Abstraction}\label{subsec:abstract}
Different from natural language, tokens in programming language are more diverse since developers can define variable names and function names in variant ways. The diversity of identifies and literals in the code
leads to more serious OOV problem during program  comprehension. Thus,
following Ahmed et al.\cite{ahmed2018compilation} and Tufano et al.\cite{tufano2019empirical}'s good practice, we adopt code abstraction to reduce vocabulary size and mitigate the OOV problem. 
\par

\begin{figure}
    \centering
    \includegraphics[scale=0.6]{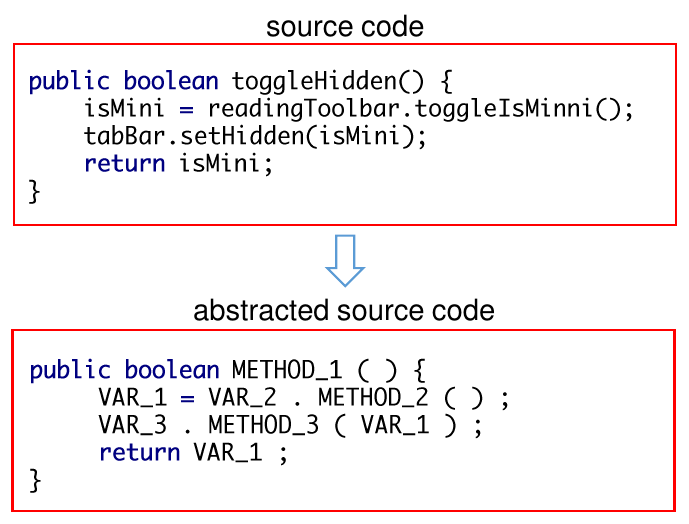}
    \caption{Example of code abstraction}
    \label{fig:abstracion}
\end{figure}

An example of code abstraction is shown in Figure~\ref{fig:abstracion}. Specifically, we use \texttt{src2abs} provided by \cite{tufano2019empirical,tufano2019learning} to abstract source code. It feeds sequence of source code to a Java parser\cite{parser} which can recognize identifiers and literals, and then generate and substitute a unique ID for each identifier and literal. If the identifier or literal appears multiple times in the same source code, it will be replaced with the same ID. Since some identifiers and literals appear frequently in the source code, they can be treated as keywords of the dataset\cite{tufano2019empirical}. The frequent identifiers and literals should not be abstracted but regarded as idioms that src2abs has provided for us.


\subsection{Dynamically Relative Position Representations}
\label{sec:dynamically}
\textbf{Relation-aware Self-Attention.} Using different position embeddings for different positions helps Transformer capture the position information of input words. However, absolute positional encoding in the vanilla Transformer is ineffective to capture the relative word orders~\cite{shaw2018self}. To encode the pairwise positional relationships between input elements, Shaw et al.\cite{shaw2018self} propose the relative position encoding which models the relation of two elements through their distance in the input sequence. Formally, the relative position embedding between input element $x_i$ and $x_j$ is represented as $a_{ij}^V$,$a_{ij}^K$ $\in$ $R^{d}$.\par
In this way, the self attention calculated in Equ. (\ref{eq:1}) and Equ. (\ref{eq:3}) can be rewritten as:
\begin{equation}
    e_{ij}=\frac{(x_iW^Q)(x_jW^K+a_{ij}^K)^T}{\sqrt{d_z}}, \label{eq:10}
\end{equation}

\begin{equation}
   z_i=\sum\limits_{j=1}^{n}\alpha_{ij}(x_{j}W^{V}+a_{ij}^V). \label{eq:9}
\end{equation}
\qquad Shaw et al.\cite{shaw2018self} also clip the maximum relative position to a maximum absolute value of $k$ since they hypothesize that precise relative position information is not useful beyond a certain distance. Clipping the maximum distance also enables the model to generalize to sequence lengths unseen during training:

\begin{equation}
    a_{i,j}^K=w_{clip(j-i,k)}^K,
\end{equation}

\begin{equation}
    a_{i,j}^V=w_{clip(j-i,k)}^V,
\end{equation}

\begin{equation}
    clip(x,k)=max(-k,min(k,x)).
\end{equation}
\qquad Hence, we learn $2k+1$ relative position representations, i.e., $w^K=(w^K_{-k},...,w^{K}_k)$ and $w^V=(w^v_{-k},...,w^V_{k})$ where $w_i^K$,$w_i^V$ $\in$ $R^{d_{model}}$.

\begin{figure*}[htbp]
    \centering
    \includegraphics[scale=0.58]{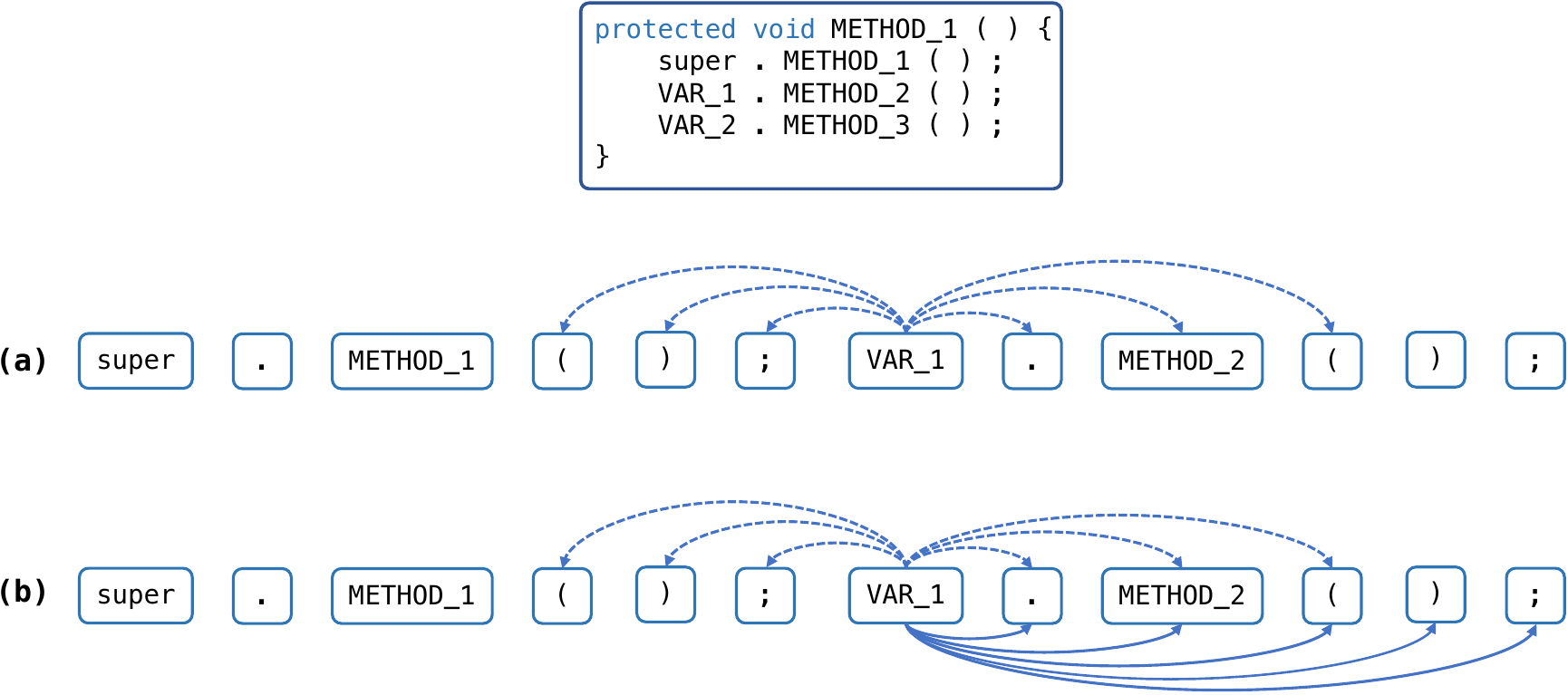}
    \caption{Illustration of the token relations for relative positions (a) and dynamically relative positions (b). For each relative position, the clipping distance $k$ is assumed as 3. Only the second and third statements of the source code are illustrated here for simplicity. Solid lines and dotted lines indicate different token relations. The dotted line represents the relative distance is smaller than $k$ and the solid line represents the tokens in the same statement with \texttt{VAR\_1}. }
    \label{fig2}
\end{figure*}
\textbf{Dynamically relation position.} 
The relative position encoding~\cite{shaw2018self} captures the pairwise positional relationships between the input elements by simply fixing the maximum relative distance at $k$. To involve the local structure information of code, we propose to incorporate the statement-level information into the position encoding. Different from pre-defining a maximum clipping distance $k$, we propose to dynamically adjust the distance based on the length of the statement during the relative position encoding, named as dynamically relation position encoding.
The difference between relative position embedding and the proposed strategy is illustrated in Figure \ref{fig2}, and the clipping distance $k$ is defined as 3. For the token \texttt{VAR\_1}, the relative position encoding enhances the relationship among the tokens before and behind the token \texttt{VAR\_1}, which is indicated with dotted lines in the relative position encoding method of Figure~\ref{fig2} (a). 
We hypothesize that tokens in one statement have stronger relations with the tokens in other statements, e.g., the token \texttt{VAR\_1} tends to be weakly relevant to the token \texttt{METHOD\_1} compared with the tokens in the same statement (e.g., token \texttt{METHOD\_2}). To incorporate statement-level syntactic information into the position embedding, we propose a dynamically relation position encoding strategy.
The proposed position encoding can help Transformer pay more attention to the tokens in the same statement (denoted as the solid lines in Figure~\ref{fig2} (b)) and the tokens with a relative distance smaller than $k$ (denoted as the dotted lines in Figure~\ref{fig2} (b)). In addition, the two kinds of attention can be superimposed in our strategy. For example, the token \texttt{METHOD\_2} receives the two kinds of attention, while the last two tokens ``\texttt{)}'' and ``\texttt{;}'' do not receive the relative position attention because their relative distance to \texttt{VAR\_1} is bigger than the clipping distance $k$ ($k=3$ here). 
In decoder, the current token cannot see the token behind, so it is impossible to get the statement mask matrix. Therefore, we only use the dynamically relative position in encoder. 

\par
\begin{figure}
    \centering
    \includegraphics[scale=0.45]{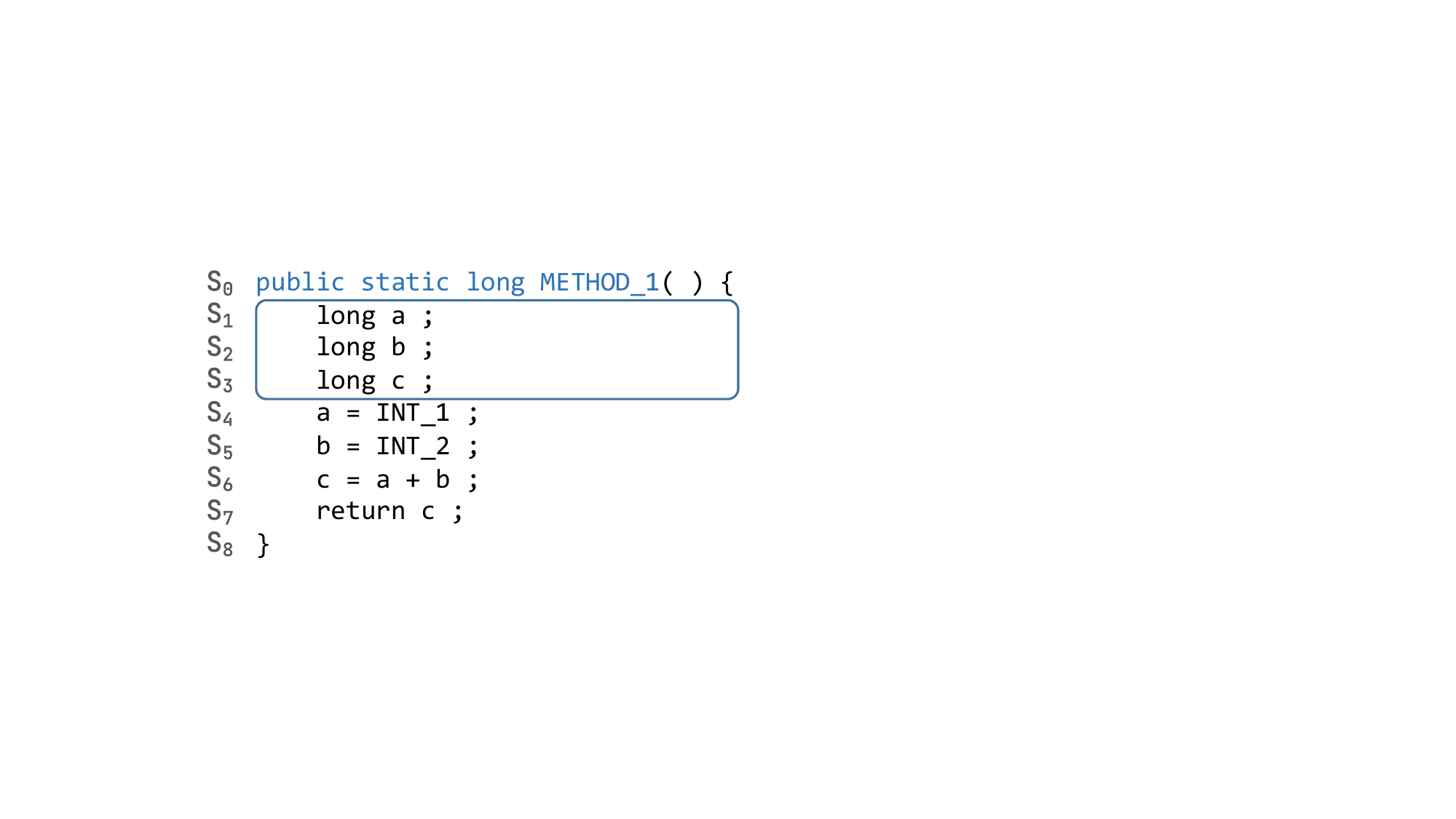}
    \caption{Example of source code to represent statement mask matrix. Note that we only take statements $s_1$, $s_2$, and $s_3$ as example for illustrating the statement mask matrix in Figure~\ref{fig:framework}.}
    \label{fig3}
\end{figure}

Similar to padding mask and sequence mask, we propose a statement mask operation to divide the code into a sequence of statements. For the code example shown in Figure \ref{fig3}, we illustrate the statement mask matrix for its statements $s_1$, $s_2$ and $s_3$ in Figure~\ref{fig:framework}. Specifically, the statement mask matrix $W^L$ is a $n\times n$ matrix which records the statement-aware information of the source code, where $n$ is the length of code tokens. For the tokens $x_i$, $x_j$ in the same statement, the value $W_{ij}^L$ between them is set as 1; otherwise the value $W_{ij}^L$ is set as 0. 

We compute the dynamically relative position embeddings as below:

\begin{equation}
  z_i=
  \begin{cases}
   \sum_{j=1}^{n}\alpha_{ij}(x_jW^V+a_{ij}^V+a^{V^{'}})   &\text{$W_{ij}^L=1$}\\
   \sum_{j=1}^{n}\alpha_{ij}(x_jW^V+a_{ij}^V)   &\text{$W_{ij}^L=0$}
  \end{cases}
  \label{eq:12}
\end{equation}
where $W^L \in R^{n\times n}$ is the statement mask matrix, and $a^{V^{'}} \in R^{d_{model}}$ is a learnable parameter vector. We then recalculate the attention weight to incorporate the dynamically relative position embeddings:
\begin{equation}
   e_{ij}=
  \begin{cases}
   \frac{x_iW^Q(x_jW^K+a_{ij}^k+a^{K^{'}})^T}{\sqrt{d_z}}   &\text{$W_{ij}^L=1$}\\
   \frac{x_iW^Q(x_jW^K+a_{ij}^k)^T}{\sqrt{d_z}}   &\text{$W_{ij}^L=0$}
  \end{cases}
  \label{eq:13}
\end{equation}

\noindent where $a^{K^{'}} \in R^{d_{model}}$ is a learnable parameter vector.

\section{Experimental Setup}\label{sec:setup}
In this section, we will introduce the benchmark datasets, implementation details, evaluation metrics and comparison models for experimentation.

\subsection{Benchmark Datasets}
We conduct evaluation on two benchmark datasets\footnote{https://sites.google.com/view/learning-codechanges}\footnote{https://sites.google.com/view/learning-fixes} following the previous work\cite{tufano2019learning,tufano2018empirical}, Gerrit\footnote{https://www.gerritcodereview.com} code reviews repository and open-source projects in GitHub (namely \gpj). 
Gerrit includes Android\cite{android}, Google Source\cite{google}, and Ovirt\cite{Ovirt}, while \gpj contains 121,895 PRs commits from GitHub open-source projects.
We classify all projects in the datasets into two levels, i.e., small level \sml and medium level \medium, according to the tokens numbers of original code. \sml and \medium contain 0-50 tokens and 500-100 tokens in each piece of original code, respectively. The two benchmark datasets are partitioned into training set (80\%), validation set (10\%) and test set (10\%) following the prior studies, with detailed statistics shown in Table~\ref{tab:benchmark}. 

\begin{table}[htbp]
  \centering
  \caption{Statistics of the two benchmark datasets.}
    \begin{tabular}{c|c|rr}
    \hline
    \hline
    \multicolumn{2}{c}{Dataset} & \sml{} & \medium{} \\
    \hline
    \multirow{4}[2]{*}{Gerrit} & Google & 2,165 & 2,286 \\
          & Android & 4,162 & 3,617 \\
          & Ovirt & 4,456 & 5,088 \\
          \cline{2-4}
          & All & 10,783 & 10,991 \\
    \hline
    \multicolumn{2}{c}{\gpj} & 58,350 & 65,545 \\
    \hline
    \hline
    \end{tabular}%
  \label{tab:benchmark}%
\end{table}%

\subsection{Implementation and Supporting Tools/Platform}

\parabf{Data Preparation.} We first abstract the source code according to Section~\ref{subsec:abstract}.
Then, we compute the statement mask matrices for the two benchmark datasets, respectively. However, computing the matrices for a large amount of code is time-consuming and inefficient, so we convert the computation into a series of matrix operations to fully use the computing resources of GPU and improve the efficiency (Section~\ref{sec:gpu} shows details). We test the time cost of the matrix computation before and after the acceleration, respectively. The results on the training set of GitProjs show that it reduces the computation time from 10 minutes to 7 seconds, indicating the efficiency of the acceleration operation.


\parabf{Hyper-parameters Setting.} \tool is composed of 6 hidden layers and 8 heads. The hidden layer size of the model and the size of every head are defined as 512 and 64, respectively. We train \tool using Adam optimizer~\cite{kingma2014adam} with an initial learning rate of $1.0$ and use warm-up~\cite{goyal2017accurate} to optimize the learning rate. We set the mini-batch size as 32 and the dropout as 0.1 during training. \tool is trained for a maximum of 20,000 steps and performed early stops if the validation performance does not improve during 2,000 steps. We also use beam search during inference and set the beam size as 10. 

\parabf{Platform.} Our experiments are conducted on a single Tesla p100 GPU for about 10 hours for \medium datasets and 5 hours for \sml datasets for both benchmark datasets, respectively.



\subsection{GPU Acceleration}
\label{sec:gpu}
Algorithm~\ref{alg1} shows how we use matrix operations to replace inefficient nested loops during computing the statement mask matrix. 

The input $X={x_1,x_2,...,x_n} \in R^n$ is the sequence of source code token. We first compute $I={i_1,i_2,...i_n} \in R^n$, which is a vector consisting of 0 and 1, from $X$. The rule for generating $I$ is : if $x_m \in X$ is an identifier, the value of $i_m \in I$ is 1; otherwise, it is 0 (Line 2). We can get $W^A \in R^n$ by multiplying $I$ and the lower triangular matrix $W^M$($W^M \in R^{n\times n}$) (Line 3-4). 
Next we will repeat $W^A$ to get $W^B \in R^{n \times n}$ (Line 5) and can find that if $i, j$ are in the same statement, $W_{i,j}^B=W_{j,i}^B$, and vice versa. So finally, if $W_{i,j}^B=W_{j,i}^B$, we let $W_{i,j}^B=W_{j,i}^B=1$; otherwise it is $W_{i,j}^B=W_{j,i}^B=0$. In this step (Line 6-10), we also use the matrix operations completely instead of nested loops, so this step is also efficient.

\begin{algorithm}
\caption{Computation of the statement mask matrix}
\label{alg1}
\begin{algorithmic}[1]
     \Require $X=(x_1,x_2,...,x_n) \in R^n$, which is a sequence of source code tokens.
   
    \Ensure the statement mask matrix, $W^L \in R^{n \times n}$
    \Function{ComputeSMM}{$X$}
    \par
    \State $ I \gets find \quad the \quad identifiers \quad from \quad X $
    \par
    // $W^M$ is lower triangular matrix, $W^M \in R^{n \times n}$
    \State $W^M \gets lower \quad triangular \quad matrix $
    \State $W^A \gets I(W^M)$
    \par
    // repeat $W^A$ $n$ times to get $W^B \in R^{n \times n}$
    \State $W^B \gets repeat(W^A)$
    \State $W^{BT} \gets (W^B)^T$
    \State $W^{S1}=|W^B-W^{BT}-1|-|W^B-W^{BT}|$
    \State $W^{S2}=|W^{BT}-W^B-1|-|W^B-W^{BT}|$
    \State $W^S=(W^{S1}+W^{S2})/2$
    \State \Return{$W^S$}
    
    \EndFunction
    
\end{algorithmic}
\end{algorithm}

\subsection{Evaluation Metrics}
We evaluate the performance of \tool in code editting using three popular metrics, including Exact Match \cite{tufano2019empirical,tufano2019learning}, BLEU-4~\cite{papineni2002bleu} and ROUGE-L \cite{lin2004rouge}. 

\textbf{Exact Match} computes the number and percentage of predicted code changes that exactly match the changed code in the test sets.

\textbf{BLEU-4} is a widely-used metric in natural language processing and software engineering fields to evaluate the quality of generated texts, e.g., machine translation, code comment generation, and code commit message generation~\cite{papineni2002bleu,DBLP:journals/corr/abs-1912-02972,DBLP:journals/corr/abs-2007-06934}. It computes the frequencies of the co-occurrence of n-grams between the ground truth $\hat{y}$ and the generated sequence $y$ to judge the similarity:
$$
\mathrm{\text{BLEU-N}}=\mathrm{b(y,\hat{y})} \cdot \exp \left(\sum_{n=1}^{N} \beta_{n} \log p_{n}(y,\hat{y})\right),
$$
where $\mathrm{b(y,\hat{y})}$ indicates the brevity penalty, and $p_{n}(y,\hat{y})$ and $\beta_{n}$ represent the geometric average of the modified n-gram precision and the weighting parameter, respectively. We use corpus-level BLEU-4, i.e., $\text{N}=4$ for evaluation since it is demonstrated to be more correlated with human judegments than other evaluation metrics~\cite{DBLP:journals/corr/abs-2010-06301}.

\textbf{ROUGE-L} is commonly used in natural language translation~\cite{lin2004rouge}, and is a F-measure based on the Longest Common Subsequence (LCS) between candidate and target sequences, where the LCS is a set of words appearing in the two sequences in the same order. 
$$
ROUGE\text{-}L=\frac{\left(1+\beta^{2}\right) R_{l c s} P_{l c s}}{R_{l c s}+\beta^{2} P_{l c s}},
$$
where $R_{l c s}=\frac{L C S(X, Y)}{len(Y)}$ and $P_{l c s}=\frac{L C S(X, Y)}{len(X)}$. $X$ and $Y$ denote candidate sequence and reference sequence, respectively. $L C S(X, Y)$ represents the length of the longest common sub-sequence between $X$ and $Y$.

\subsection{Comparison Model}

We compare \tool with three baseline models, including Tufano el al. (an NMT-based model)\cite{tufano2019learning,tufano2019empirical}, SequenceR \cite{chen2019sequencer} and CODIT\cite{chakraborty2020codit}.
Tufano el al.\cite{tufano2019empirical,tufano2019learning} employ a typical encoder-decoder model LSTM to edit method-level code, where the input is a sequence of code tokens.
SequenceR \cite{chen2019sequencer} is also LSTM-based encoder-decoder model, but it uses copy mechanism to copy code tokens from the source code during decoding. The input of SequenceR is also code token sequence.
CODIT\cite{chakraborty2020codit} is a tree-based model, which uses the ASTs of source code as input and predicts code edit at the AST level.




\begin{table*}[t]
\centering
\caption{Comparison results of \tool and token-based baselines. The \textbf{bold} indicates the best results.}\label{tab:results1}
\scalebox{1.0}{
\begin{tabular}{l|l|l|ccc|ccc}
\hline
\hline
\multicolumn{2}{c|}{\multirow{2}{*}{\textbf{Dataset}}} & \multirow{2}{*}{\textbf{Approach}} &
\multicolumn{3}{c|}{\textbf{\sml{}}} & \multicolumn{3}{c}{\textbf{\medium{}}} \\

\cline{4-9} \multicolumn{2}{c|}{}& & Exact Match & BLEU-4 & ROUGE-L & Exact Match & BLEU-4 & ROUGE-L\\
\hline
\multirow{13}{*}{\textbf{Gerrit}}

& \multirow{3}{*}{\textbf{Google}} 
& Tufano et al. &20/216(9.25\%) &55.29\% &83.81\% &17/228(7.45\%) &75.12\% &91.46\% \\

& & SequenceR & 55/216(25.46\%) & 76.87\% & 93.13\% & 43/228(18.85\%) & 89.35\% & 96.85\% \\

& & DTrans &\textbf{58/216(26.85\%)} &\textbf{78.09\%} &\textbf{93.30\%} &\textbf{63/228(27.63\%)} &\textbf{91.67\%} &\textbf{97.49\%} \\ 
\cline{3-9}
& \multirow{3}{*}{\textbf{Android}} 
&  NMT-based &79/416(18.99\%) &64.29\% &88.16\% &76/361(21.05\%) &87.33\% &96.14\% \\

& & Tufano et al. &157/416(37.74\%) & 83.86\% & 95.64\% & 83/361(22.99\%) & 90.67\% & 97.48\% \\

& & DTrans &\textbf{174/416(41.82\%)} &\textbf{84.63\%} &\textbf{95.64\%} &\textbf{112/361(31.02\%)} &\textbf{91.80\%} &\textbf{97.81\%} \\ 

\cline{3-9}
& \multirow{3}{*}{\textbf{Ovirt}} &  Tufano et al. &113/445(25.39\%) &73.60\% &91.14\% &102/509(20.03\%) &82.66\% &94.07\% \\

& & SequenceR & 173/445(38.87\%) & 85.10\% & 95.79\% & 167/509(32.80\%) & 91.69\% & 97.52\% \\

& & DTrans &\textbf{204/445(45.84\%)} &\textbf{86.81\%} &\textbf{95.81\%} &\textbf{210/509(41.25\%)} &\textbf{92.82\%} &\textbf{97.60\%} \\ 

\cline{2-9}
&\multirow{2}{*}{\textbf{Overall}} &  Tufano et al. &388/1,077(36.02\%) &82.47\% &94.57\% &334/1,098(30.41\%) &91.57\% &97.53\% \\

& & SequenceR & 405/1,077(37.60\%) & 85.82\% & 95.69\% & 284/1,098(25.86\%) & 92.04\% & 97.57\% \\

& & DTrans &\textbf{489/1,077(45.40\%)} &\textbf{86.82\%} &\textbf{96.10\%} &\textbf{409/1,098(37.24\%)} &\textbf{92.79\%} &\textbf{97.85\%} \\
\hline

\multicolumn{2}{c|}{\multirow{3}*{\textbf{\gpj{}}}}
&{Tufano et al.} & 2,119/5,835(36.31\%) & 85.84\% & 96.06\% & 1,166/6,545(17.82\%) & 90.97\% & 97.58\% \\

\multicolumn{2}{c|}{} & SequenceR & 2,255/5,835(38.64\%) & 86.72\% & 96.33\% & 1,214/6,545(18.54\%) & 91.03\% & 97.62\% \\

\multicolumn{2}{c|}{} &{\tool} & \textbf{2,573/5,835(44.09\%)} & \textbf{87.14\%} & \textbf{96.55\%} & \textbf{1,625/6,545(24.82\%)} & \textbf{91.56\%} & \textbf{97.81\%} \\

\hline
\hline
\end{tabular}
}
\end{table*}

\section{Experiment Results}\label{sec:results}
In this section, we aim at verifying the effectiveness of the proposed approach, specifically by answering the following research questions:

   
   
   

   
   

   

\begin{enumerate}[label=\bfseries RQ\arabic*:,leftmargin=.5in]
   \item What is the performance of the proposed approach compared with the baseline models?
   
   \item What is the impact of the proposed dynamically relative position encoding on the model performance?

   \item What is the effectiveness of \tool in generating multi-lines code change prediction?

   \item Whether \tool can accurately locate the lines to edit for code change prediction?

   \item What is the impact of different parameters on the model performance?
   
   \item What is the performance of \tool in cross-project setting?
   
\end{enumerate}\textbf{}

Specifically, RQ1 is to evaluate the performance of the proposed model compared with baselines, including token-based models and tree-based models. 
To verify the advantage of the proposed dynamically relative position embedding, we compare DTrans with  Transformer\cite{vaswani2017attention} and Transformer with relative position embedding (namely \rltrans)\cite{shaw2018self} in RQ2. For RQ3, since we find that more than 30\% of the code samples in the datasets need multi-line code changes, the research question is to evaluate the capacity of DTrans for generating multiple-line code changes. RQ4 is to validate the ability of locating lines to edit. Finally, since the hyper-parameters can impact the performance of DTrans, RQ5 discusses the hyper-parameter configurations. RQ6 is to evaluate the performance of \tool in cross-project setting.


\subsection{Answer to RQ1: Performance of the proposed \tool}
\label{sec:rq1}
\subsubsection{Comparison with token-based models}

Table~\ref{tab:results1} presents the experimental results of our proposed model and the token-based baselines on the benchmark datasets. 
From the table, we can observe that \tool performs better than the 
token-based baselines in predicting exact-matched code changes for all the datasets. 
For example, \tool successfully generates 489 exact-matched code changes in Gerrit for \sml and 409 for \medium, while Tufano et al. only generates 388 and 334 exact-matched code changes,  respectively, and SequenceR only generates 405 and 284 exact-matched code changes for \sml and \medium. Compared with Tufano et al. , \tool outperforms 26.04\% and 22.45\% for \sml and \medium, respectively. Compared with SequenceR, \tool outperforms 20.74\% and 44.01\% for \sml and \medium respectively.
For \gpj, SequenceR outputs 2,255 code changes that are consistent with the ground truth for \sml and 1,214 for \medium, while \tool successfully produces 2,573 and 1,625 for the two types of datasets, respectively. 
Besides, the ground truth is human-writing code \cite{tufano2018empirical,tufano2019learning}, so the higher scores of BLEU-4 and ROUGE-L represent that the results generated by \tool are semantically similar to the human-writing code.
For example, \tool increases the performance of SequenceR by 2.59\% and 0.66\% in Google \medium with respect to the BLEU-4 and ROUGE-L metrics, respectively. The results demonstrate the effectiveness of the proposed \tool over the token-based models.


\begin{table*}[t]
\centering
\caption{Comparison results of \tool, Transformer and \rltrans. The \textbf{bold} indicates the best results. ``*" denotes statistical significance in comparison to the baselines(i.e.,two-sided $t$-test with $p$-value$<$0.05)}\label{tab:results2}
\scalebox{1.0}{
\begin{tabular}{l|l|l|ccc|ccc}
\hline
\hline
\multicolumn{2}{c|}{\multirow{2}{*}{\textbf{Dataset}}} & \multirow{2}{*}{\textbf{Approach}} &
\multicolumn{3}{c|}{\textbf{\sml{}}} & \multicolumn{3}{c}{\textbf{\medium{}}} \\

\cline{4-9} \multicolumn{2}{c|}{}& & Exact Match & BLEU-4 & ROUGE-L & Exact Match & BLEU-4 & ROUGE-L\\
\hline
\multirow{12}{*}{\textbf{Gerrit}}

& \multirow{3}{*}{\textbf{Google}}
 & Transformer &40/216(18.52\%)$^*$ &73.20\%$^*$ &91.89\%$^*$ &38/228(16.66\%)$^*$ &88.49\%$^*$ &97.11\%$^*$ \\
& & \rltrans{} &58/216(26.85\%) &78.06\% &93.04\% &61/228(26.75\%) &91.08\% &97.38\% \\
& & DTrans &\textbf{58/216(26.85\%)} &\textbf{78.09\%} &\textbf{93.30\%} &\textbf{63/228(27.63\%)} &\textbf{91.67\%} &\textbf{97.49\%} \\ 
\cline{3-9}
&\multirow{3}{*}{\textbf{Andriod}}
& Transformer &146/416(35.09\%)$^*$ &83.00\%$^*$ &95.29\% &97/361(26.86\%)$^*$ &91.42\% &97.77\% \\
& & \rltrans{} &173/416(41.58\%) &84.37\% &95.58\% &99/361(27.42\%)$^*$ &91.66\% &97.62\% \\
& & DTrans &\textbf{174/416(41.82\%)} &\textbf{84.63\%} &\textbf{95.64\%} &\textbf{112/361(31.02\%)} &\textbf{91.80\%} &\textbf{97.81\%} \\ 
\cline{3-9}
&\multirow{3}{*}{\textbf{Ovirt}}& Transformer &182/445(40.89\%)$^*$ &83.73\%$^*$ &94.84\%$^*$ &172/509(33.79\%)$^*$ &92.16\%$^*$ &97.42\%$^*$ \\
& & \rltrans{} &189/445(42.47\%)$^*$ &84.82\%$^*$ &95.29\%$^*$ &188/509(36.93\%)$^*$ &92.56\% &97.55\% \\
& & DTrans &\textbf{204/445(45.84\%)} &\textbf{86.81\%} &\textbf{95.81\%} &\textbf{210/509(41.25\%)} &\textbf{92.82\%} &\textbf{97.60\%} \\ 
\cline{2-9}
&\multirow{3}{*}{\textbf{Overall}}& Transformer &428/1,077(39.74\%)$^*$ &84.37\%$^*$ &95.35\%$^*$ &355/1,098(32.33\%)$^*$ &92.19\%$^*$ &97.70\%$^*$ \\
& & \rltrans{} &472/1,077(43.82\%) &86.18\% &95.95\% &388/1,098(35.33\%) &92.29\%$^*$ &97.72\%$^*$ \\
& & DTrans &\textbf{489/1,077(45.40\%)} &\textbf{86.82\%} &\textbf{96.10\%} &\textbf{409/1,098(37.24\%)} &\textbf{92.79\%} &\textbf{97.85\%} \\
\hline

\multicolumn{2}{c|}{\multirow{3}{*}{\textbf{\gpj{}}}}
&{Transformer} & 2,503/5,835(42.89\%)$^*$ & 86.49\%$^*$ & 96.40\%$^*$ & 1,509/6,545(23.05\%)$^*$ & 91.29\%$^*$ & 97.73\%$^*$ \\
 
\multicolumn{2}{c|}{} & {\rltrans{}} & 2,540/5,835(43.53\%) & 87.01\% & 96.47\% & 1,574/6,545(24.04\%)$^*$   & 91.56\% & 97.79\% \\

\multicolumn{2}{c|}{} &{\tool} & \textbf{2,573/5,835(44.09\%)} & \textbf{87.14\%} & \textbf{96.55\%} & \textbf{1,625/6,545(24.82\%)} & \textbf{91.56\%} & \textbf{97.81\%} \\
\hline
\hline
\end{tabular}
}
\end{table*}

\subsubsection{Comparison with tree-based models}
Because CODIT does not provide the source code for data processing, and the data processing process of CODIT is very complex, we directly compare it on the code change dataset used by CODIT.
Table~\ref{tab:results3} shows the experimental results of \tool and CODIT. In the abstracted code change dataset provided by CODIT, the result of CODIT is not good. 
Compared with SequenceR, CODIT is lower than SequenceR by 17.80\%, 4.59\%, and 3.41\% with respect to the Exact Match, BLEU-4 and ROUGE-L metrics, respectively. \tool improves the performance of SequenceR by 16.10\%, 2.38\% and 0.83\% regarding the three metrics respectively.




\begin{table}[ht]
\centering
\caption{Comparison results of \tool and tree-based baseline CODIT. The \textbf{bold} indicates the best results. ``*" denotes statistical significance in comparison to the baselines(i.e.,two-sided $t$-test with $p$-value$<$0.05)}\label{tab:results3}

\begin{tabular}{c|ccc}
\hline
\hline
\textbf{Method} & Exact Match & BLEU-4& ROUGE-L\\
\hline
Tufano et al. & 1,898/5,143(36.90\%)$^*$ & 75.54\%$^*$ & 93.03\%$^*$ \\
SequenceR & 2,130/5,143(41.41\%)$^*$ & 78.02\%$^*$  & 93.87\%$^*$ \\
CODIT & 1,808/5,143(35.15\%)$^*$ & 74.59\%$^*$ & 90.77\%$^*$ \\
\hline
Transformer & 2,293/5,143(44.58\%)$^*$ & 77.87\%$^*$ & 93.98\%$^*$ \\
\rltrans & 2,426/5,143(47.17\%)$^*$ & 79.31\%$^*$ & 94.56\% \\
\hline
\tool & \textbf{2,473/5,143(48.08\%)} & \textbf{79.88\%} & \textbf{94.65\%} \\
\hline
\hline
\end{tabular}
\end{table}

\begin{tcolorbox}[breakable,width=\linewidth,boxrule=0pt,top=1pt, bottom=1pt, left=1pt,right=1pt, colback=gray!20,colframe=gray!20]
\textbf{Answer to RQ1:} In summary, \tool can more accurately predict code changes, and the generated code changes are more semantically relevant to the ground truth.
\end{tcolorbox}


\subsection{Answer to RQ2: Impact of the proposed dynamically relative position encoding on the model performance}

To evaluate the effectiveness of the proposed dynamically relative position encoding strategy, we compare \tool with the original Transformer~\cite{vaswani2017attention} and Transformer with relative position (\rltrans)\cite{shaw2018self}. 
We reproduce their experiments under the same hyper-parameter settings as \tool for fair comparison.

Table~\ref{tab:results2} shows the experimental results, we find that Transformer performs better than Tufano et al.. In more details, Transformer improves the performance of Tufano et al. by 6.3\%-123.62\%, 0.35\%-32.39\%, 0.15\%-9.64\% regarding the three metrics on the two benchmark datasets, respectively. 
Besides, Transformer can generate 4,795 code changes that exactly match the ground truth on the two benchmark datasets, which outperforms 15.32\% than SequenceR. The experimental results suggest that Transformer-based models can predict more effective code edits than token-based models.
Moreover, \rltrans{} performs better than the vanilla Transformer in most cases, which indicates that the relative position encoding in Transformer is more effective in capturing the code edit patterns. Finally, \tool achieves better performance than \rltrans{}, with increase rates at 1.28\% and 3.24\% in terms of the exact match metric on the \sml and \medium datasets, respectively. The results indicate the efficacy of the proposed dynamically relative position encoding strategy.


\begin{tcolorbox}[breakable,width=\linewidth,boxrule=0pt,top=1pt, bottom=1pt, left=1pt,right=1pt, colback=gray!20,colframe=gray!20]
\textbf{Answer to RQ2:} In summary, the Transformer-based models outperform baselines. The statement-level syntactic information for position encoding facilitates more accurate code change prediction.
\end{tcolorbox}


\subsection{Answer to RQ3: Effectiveness of code change prediction for multiple lines}



\begin{figure*}[htbp]
    \centering
    \begin{subfigure}[b]{\textwidth}
        \includegraphics[width=\textwidth]{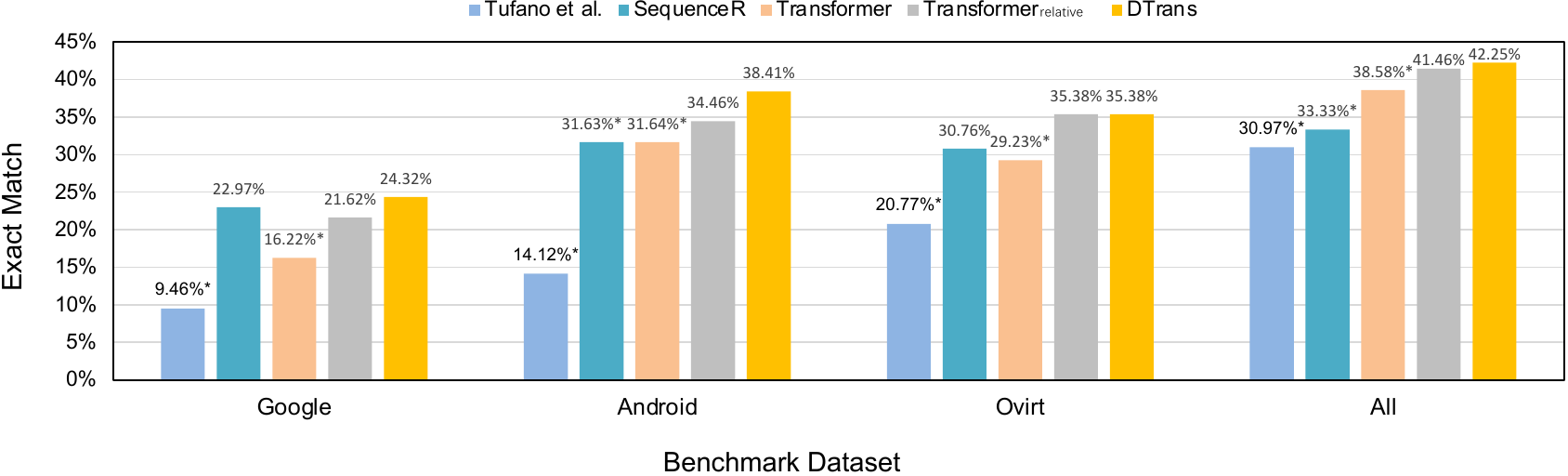}
        \caption{\sml{} projects}
      \end{subfigure}
      \hfill
      \begin{subfigure}[b]{\textwidth}
        \includegraphics[width=\textwidth]{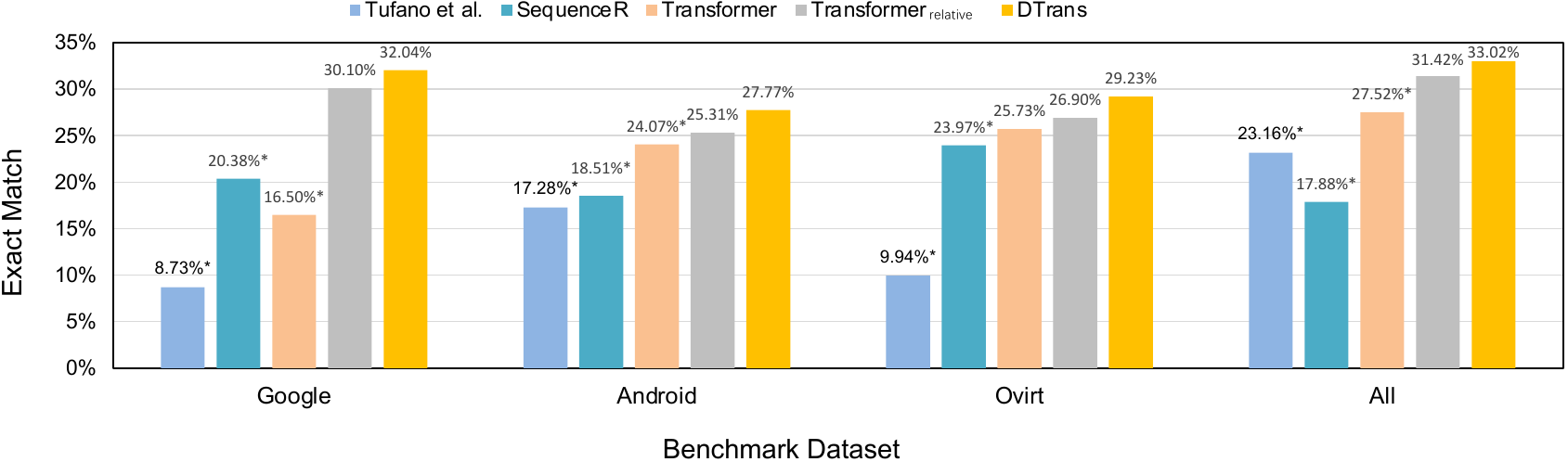}
        \caption{\medium{} projects}
        \label{fig:layer}
      \end{subfigure}
    \caption{Comparison results of generating multi-lines patch on the \sml{} (a) and \medium{} (b) projects. ``*" denotes statistical significance in comparison to the baselines (i.e., two-sided $t$-test with $p$-value$<$0.05).}
    \label{fig:multiple}
\end{figure*}

According to the statistics~\cite{chen2019sequencer,ahmed2018compilation}, most edits are accomplished through a single-line code change. However, some edits still need multi-line code changes in practice. We analyze the multi-line code changes in the Gerrit dataset~\cite{Gerrit} in Table~\ref{tab:multipleline}, and observe that nearly 35\% edits involve changes of more than one line of code.

\begin{table}[htbp]
  \centering
  \caption{Statistics of the edits that need multi-line code changes.}
    \begin{tabular}{c|ccc}
    \hline
    \hline
    \multicolumn{2}{c}{Dataset} & \sml{} & \medium{} \\
    \hline
    \multirow{4}[2]{*}{Gerrit} & Google & 74/216(34.26\%) & 103/228(45.18\%) \\
          & Android & 177/416(42.55\%) & 162/361(44.88\%) \\
          & Ovirt & 130/445(29.21\%) & 171/509(33.60\%) \\
          \cline{2-4}
          & Overall & 381/1,077(35.38\%) & 436/1,098(39.71\%) \\
    \hline
    \hline
    \end{tabular}%
  \label{tab:multipleline}%
\end{table}%

We then investigate the effectiveness of \tool in producing multi-line code changes, with the evaluation results shown in Figure~\ref{fig:multiple}. As illustrated in the table, \tool achieves the best performance among all the baselines in multi-line code change prediction.
For example, \tool overall produces 42.25\% exact-matched code changes for \sml projects and 33.02\% for \medium projects, while Tufano et al. only outputs 30.97\% and 23.16\% for the two types of datasets, respectively. The results indicate the usefulness of \tool in multi-line code prediction. We can also observe that compared with Tufano et al., the improvement of \tool on the \medium projects (42.57\%) is more significant than that on the \sml projects (36.42\%). We then analyze the average lines of code in the test sets of the \sml and \medium projects, with the results shown in Table~\ref{tab:line_number}. We can find that the code in the \medium projects are longer than that in the \sml projects on average. Therefore, we suppose the significant performance of \tool on the \medium projects may be attributed to that \tool is more effective for predicting the changes of long code snippets than baseline models.
\begin{table}[htbp]
  \centering
  \caption{Statistics of the average line of code in test sets of the \sml and \medium projects.}
    \begin{tabular}{c|ccc}
    \hline
    \hline
    \multicolumn{2}{c}{Dataset} & $M_{small}$ & $M_{medium}$ \\
    \hline
    \multirow{4}[2]{*}{Gerrit} & Google & 4.47 & 9.42 \\
          & Android & 4.89 & 10.31 \\
          & Ovirt & 4.41 & 9.01 \\
          \cline{2-4}
          & Overall & 4.60 & 9.52 \\
    \hline
    \hline
    \end{tabular}%
  \label{tab:line_number}%
\end{table}%

\begin{tcolorbox}[breakable,width=\linewidth,boxrule=0pt,top=1pt, bottom=1pt, left=1pt,right=1pt, colback=gray!20,colframe=gray!20]
\textbf{Answer to RQ3:} In summary. \tool demonstrates the superior ability of accurately generating multiple-line code changes, and has a great improvement over the baselines.
\end{tcolorbox}

\subsection{Answer to RQ4: Accuracy of \tool in locating lines to edit for predicting code change}

Locating correct lines to edit is the premise of the accurate code changes in the subsequent step. So in this research question, we analyze whether the proposed approach can accurately predict which lines to edit. 

Table~\ref{tab:results4} shows the experimental results of locating the lines for editing. 
We can observe that \tool performs better than other techniques on all projects. For example, SequenceR can only locate 66.66\% correct lines for $M_{small}$ and 54.46\% correct lines for $M_{medium}$, while \tool can locate 70.28\% and 59.01\%, respectively. This observation demonstrates that \tool can obtain more contextual information than other techniques. 

\begin{tcolorbox}[breakable,width=\linewidth,boxrule=0pt,top=1pt, bottom=1pt, left=1pt,right=1pt, colback=gray!20,colframe=gray!20]
\textbf{Answer to RQ4:} In summary, \tool can greatly outperform the baselines in locating the lines to change (e.g., achieving 8.35\% higher accuracy than the best baseline).
\end{tcolorbox}


\begin{table}[t]
\centering
\caption{Comparison results in locating lines to edit of \tool with other techniques. The \textbf{bold} fonts indicate the best results.}\label{tab:results4}
\scalebox{0.9}{
\begin{tabular}{l|l|c|c}
\hline
\multirow{1}{*}{\textbf{Dataset}} &
\multirow{1}{*}{\textbf{Approach}} &
\multicolumn{1}{c|}{\textbf{\sml{}}} & \multicolumn{1}{c}{\textbf{\medium{}}} \\

\hline
\hline
\multirow{5}{*}{\textbf{Google}}
& Tufano et al. &63/216(29.16\%) &45/228(19.73\%) \\
& Sequencer &114/216(52.77\%) &95/228(41.66\%) \\
& Transformer &91/216(42.12\%) &84/228(36.84\%) \\
& \rltrans{} &110/216(50.92\%) &118/228(51.75\%) \\
& DTrans &\textbf{116/216(53.70\%)} &\textbf{118/228(51.75\%)} \\ 
\hline
\multirow{5}{*}{\textbf{Android}}
& Tufano et al. &164/416(39.42\%) &138/361(38.22\%) \\
& Sequencer &255/416(61.29\%) &163/361(45.15\%) \\
& Transformer &235/416(56.49\%) &170/361(47.09\%)  \\
& \rltrans{} &255/416(61.29\%) &179/361(49.58\%)  \\
& DTrans &\textbf{258/416(62.01\%)} &\textbf{189/361(52.35\%) } \\ 
\hline
\multirow{5}{*}{\textbf{Overit}}
& Tufano et al. &212/445(47.60\%) &189/509(37.13\%) \\
& Sequencer &303/445(68.08\%) &311/509(61.10\%) \\
& Transformer &287/445(64.49\%) &299/509(58.74\%) \\
& \rltrans{} &302/445(67.86\%) &307/509(60.31\%) \\
& DTrans &\textbf{315/445(70.78\%)} &\textbf{328/509(64.44\%)} \\ 
\hline
\multirow{5}{*}{\textbf{Overall}}
& Tufano et al. &666/1,077(61.83\%) &587/1,098(53.46\%) \\
& Sequencer &718/1,077(66.66\%) &598/1,098(54.46\%) \\
& Transformer &690/1,077(64.06\%) &601/1,098(54.73\%) \\
& \rltrans{} &727/1,077(67.50\%) &630/1,098(57.37\%) \\
& DTrans &\textbf{757/1,077(70.28\%)} &\textbf{648/1,098(59.01\%)} \\ 
\hline
\hline
\end{tabular}
}
\end{table}

\subsection{Answer to RQ5: Impact of the model parameters}\label{sec:parameter}

In this section, we extend our experiments with different parameters to investigate the influence of internal factors of \tool.

\begin{figure*}[htbp]
    \centering
    \begin{subfigure}[b]{0.48\textwidth}
        \includegraphics[width=\textwidth]{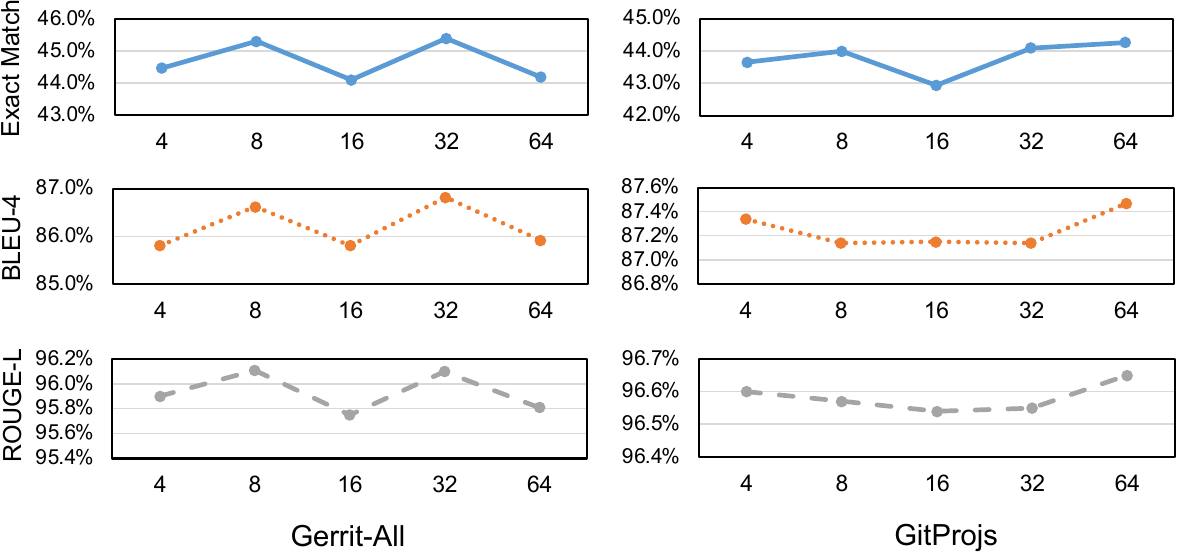}
        \caption{Clipping distance $k$.}
        \label{fig:k}
      \end{subfigure}
      \hfill
      \begin{subfigure}[b]{0.48\textwidth}
        \includegraphics[width=\textwidth]{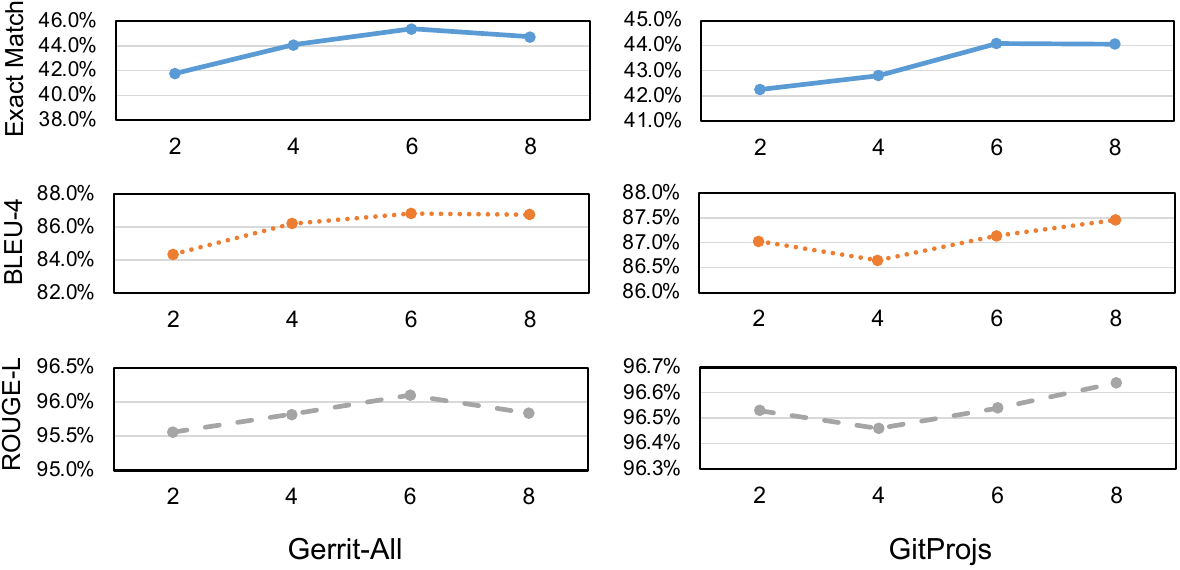}
        \caption{Number of encoder-decoder block $l$ of Transformer.}
        \label{fig:layer}
      \end{subfigure}
    \caption{Impact of key parameters on the model performance.}
    \label{fig:param}
\end{figure*}

Figure~\ref{fig:param} (a) presents the impact of the clipping distance ($k$) on the effectiveness of \tool using other default configurations (Defined in Section.\ref{sec:dynamically}). In this figure, the $x$ axis presents various clipping distances, while the $y$ axis presents the values of different evaluation metrics. We can find that  
the clipping distance does not impact the \tool effectiveness much. For example, the largest performance difference among different clipping distances is within 2\% for all evaluation metrics.
Since \tool achieves good performance on the datasets when the clipping distance is 32, we choose the parameter as 32 during experimentation.

Figure~\ref{fig:param} (b) presents the impact of the number of encoder-decoder block ($l$) on the effectiveness of \tool using other default configurations (Defined in Section.\ref{sec:transformer}). 
In this figure, the $x$ axis presents different number of encoder-decoder block, while the $y$ axis presents the values of different evaluation metrics.
From the figure, we observe that the number of encoder-decoder block has a significant impact on the model. For example, in GitProjs, the Exact Match of 2-blocks \tool is only 42.26\%, while the Exact Match of 6-blocks \tool is 44.09\%. 
Besides, more encoder-decoder blocks do not mean better performance.
For example, 6-blocks \tool is better than the 8-blocks \tool in Gerrit-All. Moreover, more encoder-decoder block will increase the model size and require more time to train. In terms of overall considerations, \tool with 6 encoder-decoder blocks is a good option.

\begin{tcolorbox}[breakable,width=\linewidth,boxrule=0pt,top=1pt, bottom=1pt, left=1pt,right=1pt, colback=gray!20,colframe=gray!20]
\textbf{Answer to RQ5:} 
In summary, the experimental results can be influenced by parameter configuration. Moreover, the clipping distance has little influence on \tool, but the number of layers has much influence on \tool.

\end{tcolorbox}

\subsection{Answer to RQ6: erformance of DTrans in cross-project setting}
In this section, we train the models in one project and test them in another project to simulate a more practical setting. We use the Gerrit dataset which contains three different projects for the evaluation. We adopt the best Transformer-based baselines for comparison. 
The results are shown in Table~\ref{tab:results_other}. We can observe that \tool consistently performs better than Transformer and \rltrans{} in the cross-project setting.
For example, when we train the
models in Google \sml and then test them in Android \sml, \tool can generate 17 exact-matched code changes, while Transformer only generates 11 exact-matched code changes. Overall, \tool increases the performance of the Transformer-based baselines by 0$\sim$200\%,0.03$\sim$5.66\%,0.09$\sim$1.30\% with respect to the Exact Match, BLEU-4, ROUGE-L metrics, respectively.
We can also find that despite the good performance of \tool in cross-project setting, it presents obvious decline compared with the in-project performance, e.g., the exact match score drops by more than 80\%.
The Transformer-based baselines show the similar trend. The phenomenon is reasonable since the edit patterns of different projects may be greatly different. 

\begin{tcolorbox}[breakable,width=\linewidth,boxrule=0pt,top=1pt, bottom=1pt, left=1pt,right=1pt, colback=gray!20,colframe=gray!20]
\textbf{Answer to RQ6:} In summary, \tool performs
better than the baseline models
in the cross-project setting. However, the performance of all the models drops greatly comparing with the in-project setting, indicating that cross-project evaluation is a more challenging setting for the code edit task.

\end{tcolorbox}

\begin{table*}[t]
\centering
\caption{Cross-project comparison results of \tool and Transformer-based baselines. The \textbf{bold} indicates the best results.}\label{tab:results_other}
\scalebox{1.0}{
\begin{tabular}{l|l|l|ccc|ccc}
\hline
\hline
\multicolumn{2}{c|}{\textbf{Dataset}} 
& \multirow{2}{*}{\textbf{Approach}} &
\multicolumn{3}{c|}{\textbf{\sml{}}} & \multicolumn{3}{c}{\textbf{\medium{}}} \\
\cline{1-2}
\cline{4-9} 
Training & Test & & Exact Match & BLEU-4 & ROUGE-L & Exact Match & BLEU-4 & ROUGE-L\\
\hline

\multirow{6}{*}{\textbf{Google}}
& \multirow{3}{*}{\textbf{Android}} 
& Transformer & 11/416(2.64\%) & 58.38\% & 86.06\% & 0/361(0\%) & 71.58\% & 90.11\% \\
&& \rltrans & 16/416(3.84\%) & 62.33\% & 87.52\% & 4/361(1.1\%) & 80.19\% & 91.82\% \\
&& \tool & \textbf{17/416(4.08\%)} & \textbf{63.92\%} & \textbf{87.74\%} & \textbf{7/361(1.93\%)} & \textbf{80.23\%} & \textbf{91.92\%} \\
\cline{2-9}
& \multirow{3}{*}{\textbf{Ovirt}}
& Transformer & 13/445(2.92\%) & 54.67\% & 81.53\% & 0/509(0\%) & 64.23\% & 83.72\% \\
&& \rltrans & 20/445(4.49\%) & 59.49\% & 82.44\% & 0/509(0\%) & 71.61\% & 84.94\% \\
&& \tool & \textbf{22/445(4.94\%)} & \textbf{61.10\%} & \textbf{82.93\%} & \textbf{1/509(0.20\%)} & \textbf{71.90\%} & \textbf{85.06\%} \\
\cline{1-9}

\multirow{6}{*}{\textbf{Android}}
& \multirow{3}{*}{\textbf{Google}} 
& Transformer & 6/216(2.77\%) & 59.66\% & 83.69\% & 2/228(0.87\%) & 73.35\% & 87.81\% \\
&& \rltrans & 8/216(3.70\%) & 62.37\% & 83.82\% & 2/228(0.87\%) & 75.30\% & 87.83\% \\
&& \tool & \textbf{10/216(4.62\%)} & \textbf{63.28\%} & \textbf{84.09\%} & \textbf{2/228(0.87\%)} & \textbf{76.15\%} & \textbf{88.26\%} \\
\cline{2-9}
& \multirow{3}{*}{\textbf{Ovirt}}
& Transformer & 21/445(4.71\%) & 61.20\% & 82.86\% & 0/509(0\%) & 68.23\% & 84.61\% \\
&& \rltrans & 22/445(4.94\%) & 64.84\% & 83.61\% & 1/509(0.19\%) & 71.42\% & 85.02\% \\
&& \tool & \textbf{30/445(6.74\%)} & \textbf{64.86\%} & \textbf{83.70\%} & \textbf{3/509(0.58\%)} & \textbf{72.05\%} & \textbf{85.27\%} \\

\cline{1-9}

\multirow{6}{*}{\textbf{Ovirt}}
& \multirow{3}{*}{\textbf{Google}} 
& Transformer & 6/216(2.77\%) & 56.78\% & 82.29\% & 0/228(0\%) & 62.90\% & 81.34\% \\
&& \rltrans & 10/216(4.62\%) & 57.99\% & 82.78\% & 1/228(0.43\%) & 65.13\% & 82.54\% \\
&& \tool & \textbf{10/216(4.62\%)} & \textbf{59.06\%} & \textbf{82.86\%} & \textbf{1/228(0.43\%)} & \textbf{68.82\%} & \textbf{83.60\%} \\
\cline{2-9}
& \multirow{3}{*}{\textbf{Android}}
& Transformer & 25/416(6.01\%) & 60.94\% & 86.70\% & 1/361(0.27\%) &72.24\% & 88.60\%\\
&& \rltrans & 23/416(5.52\%) & 62.60\% & 87.02\% & 4/361(1.10\%) & 73.38\% & 88.91\% \\
&& \tool & \textbf{25/416(6.01\%)} & \textbf{63.75\%} & \textbf{87.50\%} & \textbf{5/361(1.38\%)} & \textbf{76.64\%} & \textbf{90.07\%} \\

\hline
\hline
\end{tabular}
}
\end{table*}


\section{Case Study}\label{sec:case}

\begin{figure*}[htbp]
    \centering
    \includegraphics[width=0.8\textwidth]{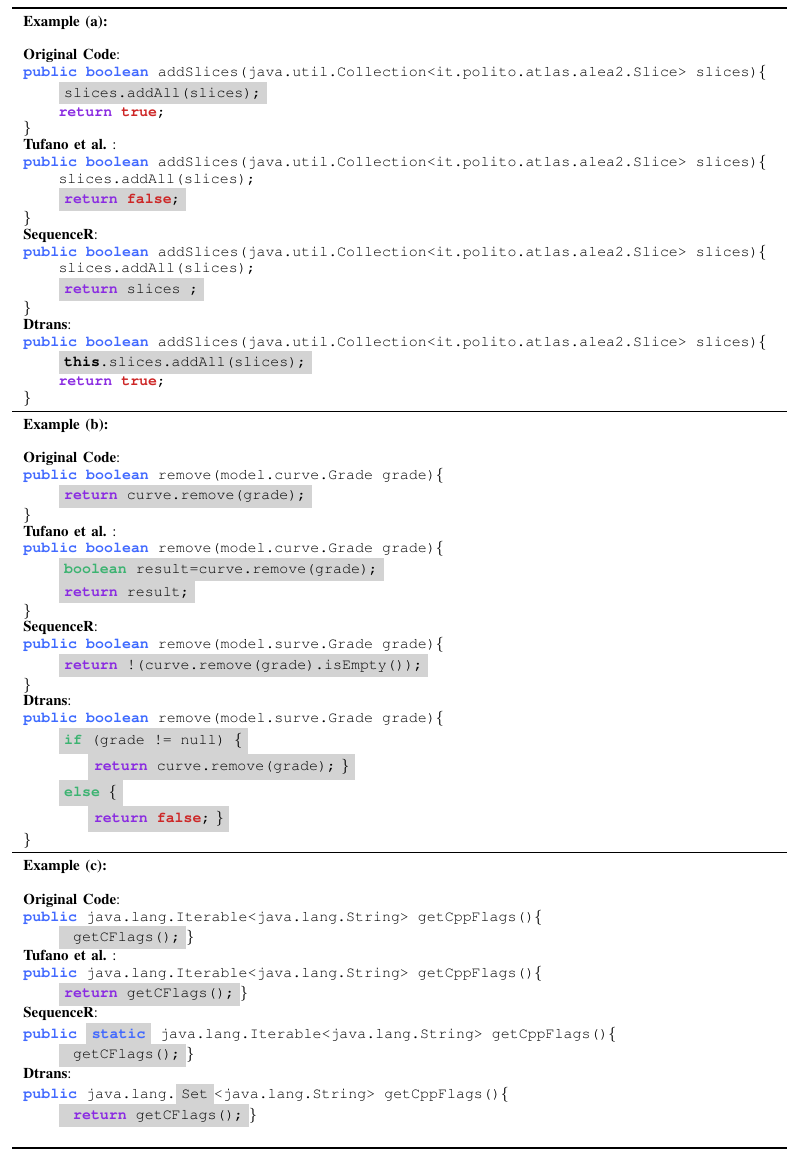}
    \caption{Qualitative analysis of our \tool model and the baseline models on three examples. }
    \label{fig:case}
\end{figure*}

To evaluate the performance of \tool in predicting accurate code edits, we select three cases from benchmark datasets as shown in Figure~\ref{fig:case}.




Figure~\ref{fig:case} (a) presents an example of code edit operation prediction. 
The method \texttt{addSlices} lacks an object to conduct the method function \texttt{slices.addAll(slices)}, so the correct edit operation is to add an object \texttt{this}.
However, Tufano et al. mistakenly predicts the operation is to change the return from \texttt{true} to \texttt{false}. Similarly, SequenceR incorrectly returns the variable \texttt{slices}. 
\tool successfully predicts the correct edit operation and adds \texttt{this} to point to the variable inside the class.

In Figure~\ref{fig:case} (b), the original code needs to remove \texttt{grade} from \texttt{curve}, but it does not check whether the variable \texttt{grade} is \texttt{null}. The correct edit operation is to check variable \texttt{grade} before executing \texttt{remove}.
Tufano et al. does not check variable \texttt{grade} and just refactors original code. 
SequenceR predicts the correct operation method but the incorrect operation object. It checks \texttt{null} for \texttt{curve.remove(grade)} rather than variable \texttt{grade}.
\tool successfully predicts both the correct operation method and operation object. It checks whether the variable \texttt{grade} is \texttt{null} before executing \texttt{curve.remove(grade)}.

In addition, \tool does not always predict accurate code changes.
In Figure~\ref{fig:case} (c), the original code needs a \texttt{return} statement because the modifier \texttt{void} does not appear in the method definition. Tufano et al. successfully adds \texttt{return} before \texttt{getCFlags()}. Sequencer mistakenly thinks that the original code should be a static function, so it inserts the modifier \texttt{static}. For \tool, it successfully adds the \texttt{return} token, but incorrectly changes the API from \texttt{java.lang.Iterable} to \texttt{java.lang.Set}, which is a non-existent interface. This example motivates us to create an API knowledge base to facilitate the code edit process in future.



\section{Threat to Validity}\label{sec:threat}

\textbf{Internal validity} is mainly about the  hyper-parameter configuration we adopted in our \tool model. To reduce this threat, we conduct an experiment to study the impact of configuration, and 
we explain in Section~\ref{sec:parameter} about how hyper-parameters influence our model's performance.

\textbf{Construct validity} is mainly the suitability of our evaluation metrics. To reduce this risk, we additionally introduce BLEU-4 (Bilingual evaluation understudy in 4-gram)~\cite{papineni2002bleu} and ROUGE-L (Recall-Oriented Understudy for Gisting Evaluation in Longest Common Sub-sequence)\cite{lin2004rouge} to evaluate the effectiveness of our approaches, which can well simulate the non-trivial coding activities in evaluating generated code.

\textbf{External validity} is mainly concerned with whether the performance of our \tool techniques can still effective in other datasets. To reduce these threats, we additional select 2.3 million pair-wise code changes generated from GitHub open-source projects\cite{tufano2019empirical} to evaluate the effectiveness of our approach. And experimental results demonstrate the effectiveness of our approach (in Section~\ref{sec:rq1}).
To further reduce the threats, we are planning to collect more open-source projects to evaluate our approach. Besides, the quality of the datasets may be another threat. In this paper, we simply follow the previous work \cite{tufano2019learning,chakraborty2020codit} by directly adopting the benchmark datasets without further cleaning. As illustrated in Sun et al.~\cite{sun2022importance}, high-quality datasets are very important for DL models. We will study the quality of the datasets in future work.
\section{Related Work}\label{sec:literature}

Related works focus on two key aspects: position representations of Transformer and automatic code edit.

\subsection{Position Representations of Transformer}

Unlike RNN\cite{schuster1997bidirectional}, which incorporates inductive bias by successively loading the input tokens, Transformer is less position-sensitive\cite{vaswani2017attention}. It is critical to incorporate position encoding into the Transformer.

\textbf{Absolute Position Representations.} Vaswani et al.\cite{vaswani2017attention} proposed Transformer and trigonometric function to calculate positional information for each token, but the positional information cannot change, while Devlin et al.\cite{devlin2018bert} and Liu et al\cite{liu2019roberta} use parameter matrix to calculate positional information.
Liu et al.\cite{liu2020learning} proposed FLOATER, which models position encoding as a continuous dynamical system and admit the standard sinusoidal position encoding as a special case, making more flexible in theory. Dehghani et al.\cite{dehghani2018universal}and Lan et al.\cite{lan2019albert} found that injecting the position information into layers can improve performance of Transformer in some tasks.

\textbf{Relative Position Representations.} Relative position Representations take the relative distance into calculating attention rather than absolute position, which performs more effective and flexible. 
Shaw et al.\cite{shaw2018self} first proposed the concept of relative position embedding and its application scope.
Yang et al. \cite{yang2019xlnet} and Dai et al.\cite{dai2019transformer} improved the relative position embedding to boost the effectiveness.
Raffel et al.\cite{raffel2019exploring} and Ke et al.\cite{ke2020rethinking} evaluated the effective of ``input-position" and ``position-input" and remove them from Transformer.
He et al.\cite{he2020deberta} evaluated the absolute and relative position embedding and proved the usability of relation position embedding. 
Since these approaches are developed for natural language processing, they are unable to capture the statement-level information included in code; while our proposed dynamically relative position encoding strategy is specifically designed for involving the statement-level syntax information of source code.

\subsection{Automatic Code Edit}
Code edit throughout the program development and maintenance relates to various behaviors, e.g., automatic program repair\cite{liu2021critical,liu2018closer,wang2021beep,wang2021syncobert,tian2020evaluating}, API-related update\cite{nguyen2010graph}, and code refactoring\cite{ge2012reconciling,raychev2013refactoring}. In recent years more and more proposed works adapted Deep Learning (DL) techniques in automatic code edit~\cite{chakraborty2020codit,boshernitsan2007aligning,meng2015does,tian2020evaluating}, aiming at automatically predicting code changes using a data-driven approach. Tufano et al. \cite{tufano2019learning} applied Neural Machine Translation (NMT) techniques to generate target code at the method level. They treated code as natural language, converting it into tokens and using code abstraction to overcome the issue of \textit{out-of-vocabulary}. Chen et al.\cite{chen2019sequencer} presented the SequenceR, an NMT-based approach, which uses the attention mechanism and outperforms Tufano et al.\cite{tufano2019learning}. Chakraborty et al.\cite{chakraborty2020codit} presented CODIT, a tree-based NMT model for predicting concrete source code changes and learning code change patterns in the wild, and it is the state-of-the-art NMT-based model in code edit.
Above approaches ignore the statement-level information, 
so we propose \tool, a novel Transformer-based approach, which explicitly incorporates the statement-level syntactic information for better capturing the local structure of code, to predict code changes.

\section{Conclusion and Future Work}\label{sec:con}
In this paper, we introduced \tool, a Transformer-based technique that can predict code changes from merged pull requests codes from developers. 
To better capture the statement-level information of code, \tool is designed with dynamically relative position encoding in multi-head attention of Transformer. 
Compared with other DL-based techniques such as neural machine translation (NMT), \tool can capture the syntactic information, which makes the generated code changes higher-quality.
The experimental results show that \tool can more accurately generate program changes in automatic code edit.

Our experiments also demonstrate the difficulties in the cross-project code edit task. In the future, we plan to investigate the cross-project challenges and incorporate more semantic information(e.g., Control-Flow Graph, Abstract Syntax Tree) to increase our capacity in code editing for the cross-project task.

\section{Acknowledgments}
This research was supported by National Natural Science Foundation of China Grant under project No. 62002084, 61872110, 61672191, Stable support plan for colleges and universities in Shenzhen under project No. GXWD2020 1230155427003-20200730101839009, the Major Key Project of PCL (Grant No. PCL2022A03, PCL2021A02, PCL2021A09), Guangdong Provincial Key Laboratory of Novel Security Intelligence Technologies (2022B1212010005), the Science and Technology Program of Guangzhou, China (202103050004).





\ifCLASSOPTIONcaptionsoff
  \newpage
\fi



%
\bibliographystyle{IEEEtran}
\bibliography{sigproc}

%








\end{document}